\documentclass{
article}

\usepackage[english]{babel}

\usepackage{amsmath}
\usepackage{graphicx}
\usepackage[colorlinks=true, allcolors=blue]{hyperref}
\usepackage{wrapfig}
\usepackage{subcaption}
\usepackage{arxiv}
\usepackage[utf8]{inputenc} 
\usepackage[T1]{fontenc}    
\usepackage{url}            
\usepackage{booktabs}       
\usepackage{amsfonts}       
\usepackage{nicefrac}       
\usepackage{microtype}      
\usepackage{lipsum}		
\usepackage{doi}
\usepackage{tabularx}
\usepackage{array}
\providecommand{\undertitle}[1]{}
\usepackage[numbers,sort&compress]{natbib}

\title{Simulating Multi-Stakeholder Decision-Making with Generative Agents
in Urban Planning}
\author{ \href{https://orcid.org/0000-0002-1595-6895}{\includegraphics[scale=0.06]{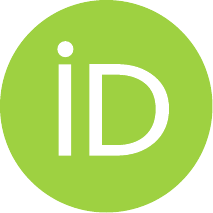}\hspace{1mm}Jin Gao}
  \\
	Massachusetts Institute of Technology\\
	Cambridge, MA 02139 \\
	\texttt{gaojin@mit.edu} \\
	\And
	\href{https://orcid.org/0009-0000-3403-7061}{\includegraphics[scale=0.06]{orcid.pdf}\hspace{1mm}Hanyong Xu} \\
	Massachusetts Institute of Technology\\
	Cambridge, MA 02139 \\
	\texttt{hanyongx@mit.edu} \\
 	\And
	\href{https://orcid.org/0009-0003-9460-7697}{\includegraphics[scale=0.06]{orcid.pdf}\hspace{1mm}Luc Dao} \\
	Massachusetts Institute of Technology\\
	Cambridge, MA 02139 \\
	\texttt{daoluc@mit.edu} \\
}

\begin{document}
\maketitle

\begin{abstract}
Reaching consensus in urban planning is a complex process often hindered by prolonged negotiations, trade-offs, power dynamics, and competing stakeholder interests, resulting in inefficiencies and inequities. Advances in large language models (LLMs), with their increasing capabilities in knowledge transfer, reasoning, and planning, have enabled the development of multi-generative agent systems, offering a promising approach to simulating discussions and interactions among diverse stakeholders on contentious topics. However, applying such systems also carries significant societal and ethical risks, including misrepresentation, privacy concerns, and biases stemming from opinion convergence among agents, hallucinations caused by insufficient or biased prompts, and the inherent limitations of the foundation models. To evaluate the influences of such factors, we incorporate varying levels of real-world survey data and demographic detail, to test agents’ performance on two decision-making value frameworks, altruism-driven and interest-driven, on a real-world urban rezoning challenge. This approach evaluates the influence of demographic factors such as race, gender, and age on collective decisions in the design of multi-generative agent systems. Our experimental results reveal that integrating demographic and life-value data enhances the diversity and stability of agent outputs. In addition, communication among the generated agents improves the quality of collective reasoning. These findings provide a predictive framework for decision-makers to anticipate stakeholder reactions, including concerns, objections, and support. By enabling the iterative refinement of proposals before public release, the simulated approach fosters more equitable, and cost-effective decisions in urban planning.
\end{abstract}


\section*{Introduction}
A large language model (LLM) is a type of machine learning model that generates human-like text by learning from vast amounts of data. Recently, LLMs have shown strong capabilities in transferring knowledge, generalizing across contexts, and performing reasoning and planning tasks. One exciting development is their integration into multi-generative agent systems, which combine distinct personas, tools, data, and memory to enable rich interactions among agents. By assigning different roles, expertise, or personality traits to each agent, these systems can simulate complex collaborations and communications.

Multi-generative agent systems are well-suited for modeling complex social and public decision-making processes. Cities, in particular, provide an ideal testing ground: urban policy discussions often involve multiple levels of government and diverse stakeholders, with processes such as proposals, hearings, lobbying, and voting. While these procedures aim to ensure that different interests are considered, they can also lead to protracted negotiations and unforeseen complications. In our study area in the US, for example, the final decision is made by a vote from a planning committee of senior officials, but only after extensive public hearings and discussion among stakeholders.

Designing such agent-based simulations for real-world decision-making carries important societal and ethical considerations. First, if the data used to train or inform these agents contains biases, the system may unintentionally reinforce those biases—amplifying inequalities rather than promoting fair outcomes. Second, when agents are modeled on real individuals, using their personal opinions to generate responses raises questions about consent and privacy. Third, an agent’s “view” depends on the data it sees and the limited context window of the underlying LLM; as a result, simulations may fail to capture the full diversity of stakeholder perspectives in a community.

This research evaluates how incorporating stakeholder perspectives—particularly life values and demographic characteristics—affects the quality and fairness of simulated conversations. Specifically, we ask: \textit{\textbf{“How do different modes of communication, agent life values, and demographic variables influence collective decisions? Should demographic attributes be included when designing such multi-generative agent systems?”}} We test these questions in the context of urban planning. In our case study, agents representing various demographic groups collaborate to propose a redevelopment plan for a rapidly growing neighborhood, using real survey data to ground their preferences and priorities.

\section{Related Works}
\subsection{Multi-Generative Agent System Applied in Decision-Making}
Building upon the foundation of traditional agent-based modeling, which relies on rule-based or stochastic agent behaviors, current research extends these systems by incorporating the reasoning, action, and memory capabilities of LLMs. This advancement enables more realistic simulations of collective decision-making. One of the early influential works is Generative Agents by Park et al.\cite{park2023generative}, which demonstrates how agents endowed with observation, planning, and reflection capabilities can autonomously make decisions within small-scale communities. To further structure these emergent behaviors, researchers have incorporated social structures and mechanism design into multi-agent systems. For example, Ren et al.\cite{ren2024emergence}developed a multi-agent framework to study the emergence of social norms, while Dai et al.\cite{dai2024artificial}examined how environmental conditions can lead agents to spontaneously form social contracts, thus fostering social norms and reducing conflicts within generative multi-agent systems. Another promising direction is enhancing decision-making quality through cross-disciplinary interactions among different expert agent groups, as shown by Costabile et al.\cite{costabile2025factchecking}, who demonstrated performance exceeding that of human crowds in fact-checking tasks. Additionally, integrating numerical strategic planning with language-based reasoning has also shown great promise. For instance, Cicero\cite{bakhtin2022diplomacy}claimed to achieve human-level performance in playing the strategic game Diplomacy.

\subsection{Real-World Data Augmented Agent-Based Simulations}
Purely generative agent-based models lack grounding in actual human data, limiting their real-world applicability; to overcome this, researchers explored integrating real-world demographic and behavioral data into generative agents to enhance simulation realism and representativeness. For instance, Park et al.\cite{park2024thousand}simulated over a thousand real individuals by conditioning LLM agents on extensive personal interviews. This approach achieved a high (85\%) accuracy in reproducing the individuals' survey responses, comparable to the consistency of participants' own responses over time. Chopra et al.\cite{chopra2024limits}and Hou et al. \cite{hou2025vaccine}modeled generative agents assigned demographic attributes derived from census data, demonstrating how demographic factors influence health-related social behaviors and decisions grounded in real-world data. Notably, Chopra’s work demonstrated the potential for very large-scale simulation with millions of agents by sampling "LLM archetypes"—representative groups of agents with similar behavior patterns—thereby allowing for nuanced behaviors without the need for individual LLMs for each agent.

\subsection{Evaluation and Mitigation of Multi-Agent Emergent Bias}
Studies revealed multi-generative agent systems can develop emergent biases due to inherent bias on training data, real-world data sample’s statistic bias, polarization effects\cite{piao2025polarization}, social conventions\cite{ashery2025conventions}or interactions between agents\cite{ranjan2025fairness}. One common strategy to evaluate the agents’ biased outputs is sentiment analysis. Elizabeth et al.\cite{ondula2024sentimental}performed automated sentiment analysis to track the tone of each agent’s utterances throughout a group discussion, to evaluate how emotional tone volatility might correlate with opinion changes. Another approach is word frequency and lexical bias analysis, Bai et al.\cite{bai2024fairmonitor}uses both human scoring and word frequency statistics to detect stereotypes and biases in a range of educational scenarios.

\section{Methodology}

\subsection{Kendall Square Redevelopment Proposal}
We chose Kendall Square, located in the heart of Cambridge, Massachusetts, USA, as our case study site. Over the past few decades, this area has undergone dramatic development, becoming a hub for startups and high-tech industries. However, the rapid growth has also raised concerns about inclusiveness and affordability, with criticism centered on the displacement caused by gentrification. In 2017, the relocation of the John A. Volpe National Transportation Systems Center left a 14-acre parcel of land available for urban redevelopment \cite{mit2021volpe}.  This decision raised a key challenge: \textbf{how to balance economic growth with social responsibility in a rapidly evolving urban landscape.} To address this, we propose two contrasting visions for the use of the site and ask the agents to discuss how much they agree or disagree with the proposals:

\textbf{Altruism-driven:} Develop low-income housing to address homelessness and the rising living costs. 

\textbf{Interest-driven:} Develop a shopping mall to create jobs and stimulate the local economy.

\begin{wrapfigure}{r}{0.45\textwidth}
  \centering
  \vspace{-0.9cm}
  \includegraphics[width=\linewidth]{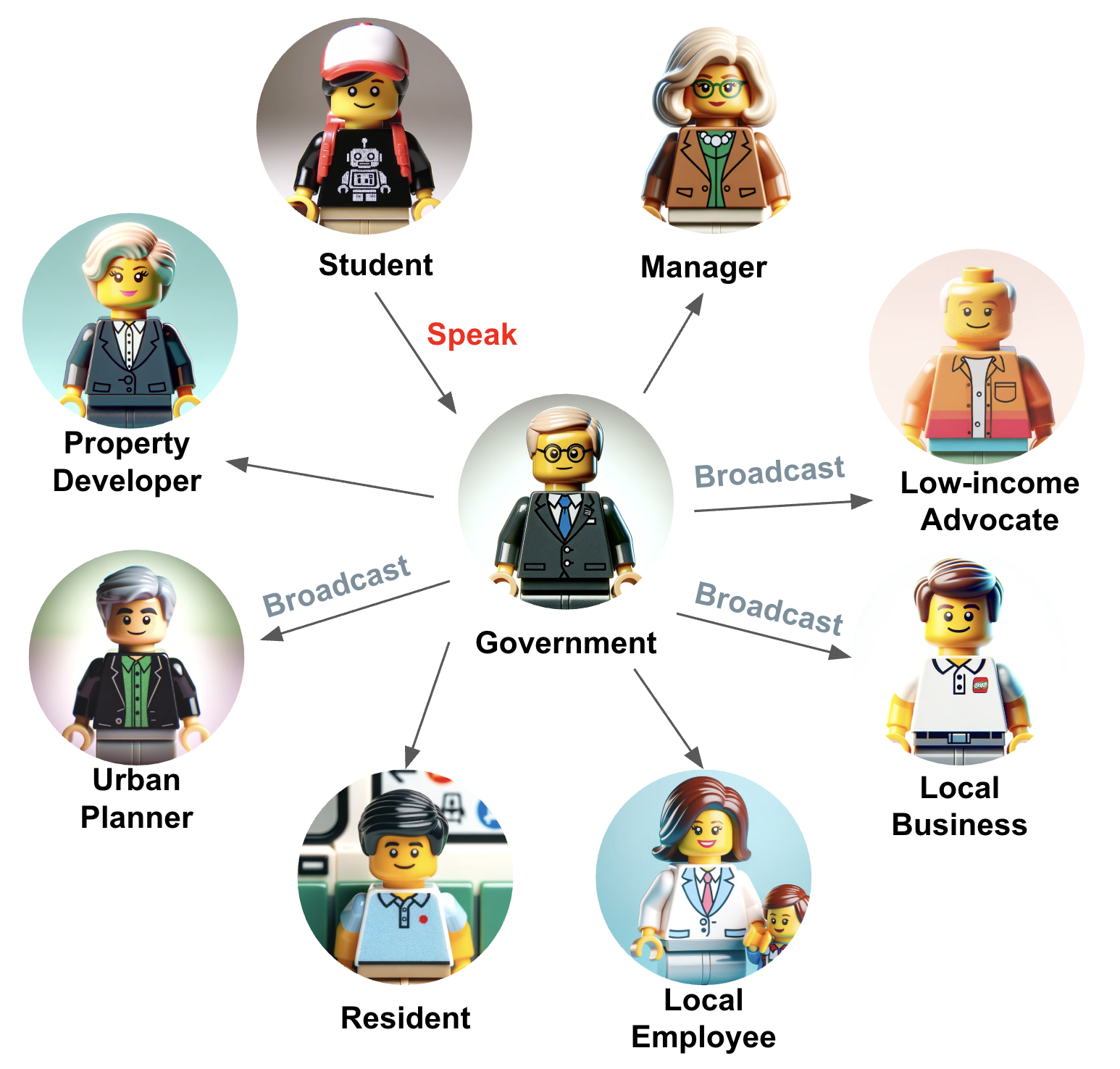}
  \caption{Agent Communication Setup}
  \label{fig:communication}
\end{wrapfigure}

\subsection{Framework and Agent Design}

We developed our experiments on AutoGen framework \cite{wu2023autogen}, which is flexible in defining agent roles,
interaction and customizations including prompts, human-in-the-loop and tool usage.

To simulate government-led community negotiations, we refactor the framework to facilitate a series of group chat
simulations with the structure shown by Figure~\ref{fig:communication}. We developed eight generative agents representing
eight different stakeholders. Each agent connects to ChatGPT-4 Turbo APIs, starting with a prompt to describe a stakeholder.
The prompt consists of four components:

\begin{itemize}
  \item \textbf{Role}: a high-level description of the stakeholder
  \item \textbf{Demographics}: demographic variables such as age, gender, race, ethnicity, etc.
  \item \textbf{Daily Life/Value}: detailed description of the stakeholder’s daily life or personal opinion based on the survey or interview
  \item \textbf{Task and Format}: ensure the agents participate in the discussion based on their own description within the ChatGPT API context window limit.
\end{itemize}

The Government is an admin agent, coordinating the group chat by proposing a topic, prompting the next agent and sharing information. Each generative agent takes turns to opine about the topic. When an agent speaks, the Government collects the message and broadcasts it to all other generative agents so that all the agents receive and process each other’s opinions. If users have inputs, the government also represents the user-agent to convey human controllers’ messages to the shared discussion context, to enable human-in-the-loop interactions.

The agent prompt is developed based on actual interviews conducted on site. The interviews include questions about the interviewees’ work, life, and their opinions about the Kendall Square Initiative, which is an urban development project in Main Street to revitalize the area and make it a new entrance into the city. Due to the context window limit of ChatGPT API, we summarize the interviews into the Role, Demographic, and Daily Life of the agents.

\subsection{Simulation Experiments}
To facilitate the discussion among the agents, we set the simulation run as the following steps: 

\begin{figure}
    \centering
    \includegraphics[width=1\linewidth]{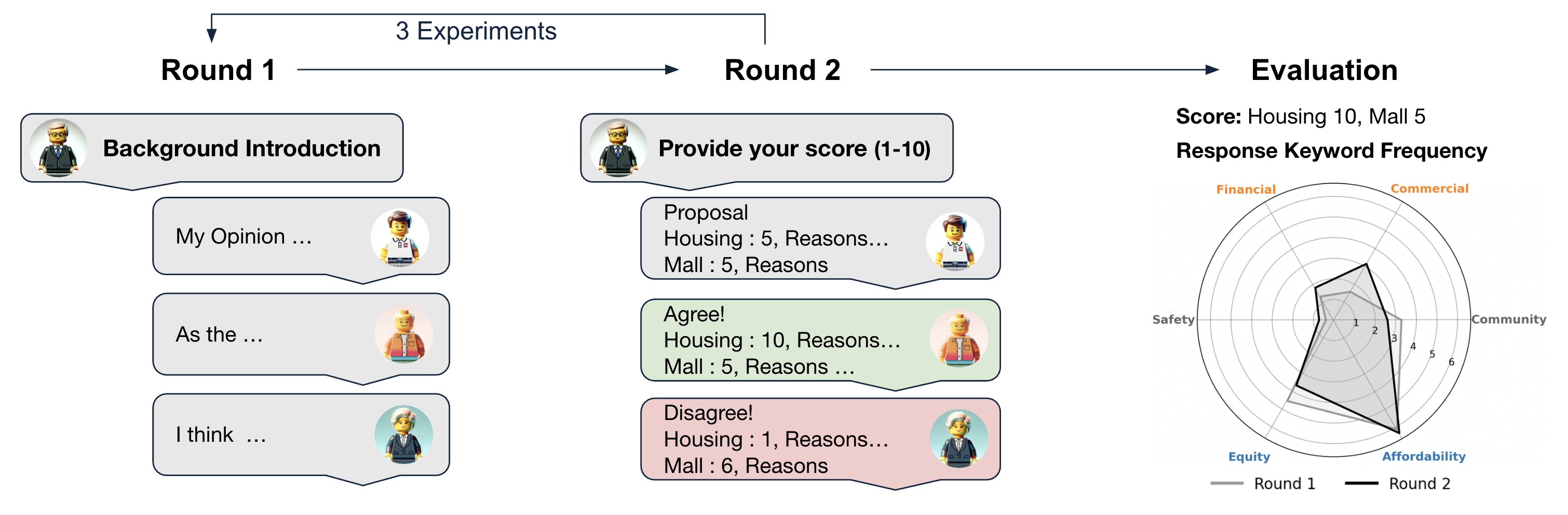}
    \caption{Agent Discussion and Evaluation Procedure}
    \label{fig:Agent Discussion and Evaluation Procedure}
\end{figure}

\begin{enumerate}
  \item The generative agents are initialized with the profile prompt.
  \item The Government starts the discussion by providing the problem context, the proposals as well as pros and cons of each proposal.
  \item The Government requests all agents to provide their opinions.
  \item When an agent speaks to the Government, the message is broadcast to all other agents to ensure all agents are aware of each other's opinions.
  \item Once all agents have spoken, the Government requests all agents to vote from 0 (disagree) to 10 (agree) for each proposal.
\end{enumerate}

We keep the temperature parameter, which decides the randomness and creativity of responses, as 1 (default value, medium creativity). The agents may generate different results each run. We repeat the run 3 times and record all results for analysis purposes. To assess how the communication, survey data, and demographic variables impact the agents’ decisions, we adjust the prompts with 6 different setups as per Table 1. For example, we compare the results of setup 2 to setup 1 to assess the impact of Life/Value on the agent response and decisions without communication.

\begin{table}[htbp]
\centering
\caption{Four different setups to assess prompt elements.}
\label{tab:prompt_setups}
\begin{tabular}{lcccccc}
\toprule
\textbf{Feature} & \textbf{Setup 1} & \textbf{Setup 2} & \textbf{Setup 3} & \textbf{Setup 4} & \textbf{Setup 5} & \textbf{Setup 6} \\
\midrule
Communication & $\times$ & $\times$ & $\times$ & \checkmark & \checkmark & \checkmark \\
Role          & \checkmark & \checkmark & \checkmark & \checkmark & \checkmark & \checkmark \\
Demographic   & $\times$ & $\times$ & \checkmark & $\times$ & $\times$ & \checkmark \\
Life/Value    & $\times$ & \checkmark & \checkmark & $\times$ & \checkmark & \checkmark \\
\bottomrule
\end{tabular}
\end{table}

\section{Findings}
\subsection{Impact of Communication}

As explained in the previous section, we compared the single agent planning decisions with the multi-agent decisions to confirm the effectiveness of communication in the multi-agent decision-making system. Below is an example of the planning decisions and excerpts of the reasoning in both scenarios of the agent University Student. We have also highlighted the main idea or ideas that are worth comparing with bold font.
Single-Agent Result: My inclination is towards the construction of low-income housing. Despite safety concerns, this option supports economic diversity, vital for a healthy, balanced community. Affordable housing addresses the pressing local need and enhances my environment by potentially reducing the number of people experiencing homelessness, thus potentially increasing safety in the long run. The strain on public services is a valid concern, but it offers an opportunity for the community to rally together to improve and scale these services to meet demand. [...]

\textit{Multi-Agent Result: From my perspective, transforming the government building into low-income housing emboldens the ethos of social responsibility and diversity that I value as part of the student community. Beyond the personal satisfaction of living within a diverse neighborhood, low-income housing could enrich the local talent pool, potentially offering unique collaboration opportunities for community-focused research or student-led initiatives with direct societal impact. The concerns over safety and the financial strain on public services are not to be dismissed lightly. However, integrating a diverse socioeconomic demographic can stimulate broader community engagement and foster innovative approaches to urban living and governance - a relevant context for a university student researching socio-technical systems. On the contrary, while the shopping mall might provide Local Business convenience and tax revenue, it is less aligned with my academic environment’s aspirations. [...]}

To interpret each stakeholder’s reasoning, we analyzed the frequency of predefined keywords related to six criteria—Safety, Affordability, Commercial, Financial, Community, and Equity—grouped into three color-coded categories: Altruistic (blue: Affordability and Equity), Neutral (gray: Community and Safety), and Interest-Driven (orange: Financial and Commercial). Specifically, for each stakeholder role and experimental setup, we calculated the average keyword frequency per turn by averaging the total keyword occurrences across three repetitions. The results are illustrated by the series of radar charts in Figure 3.

\begin{figure}
    \centering
    \includegraphics[width=1\linewidth]{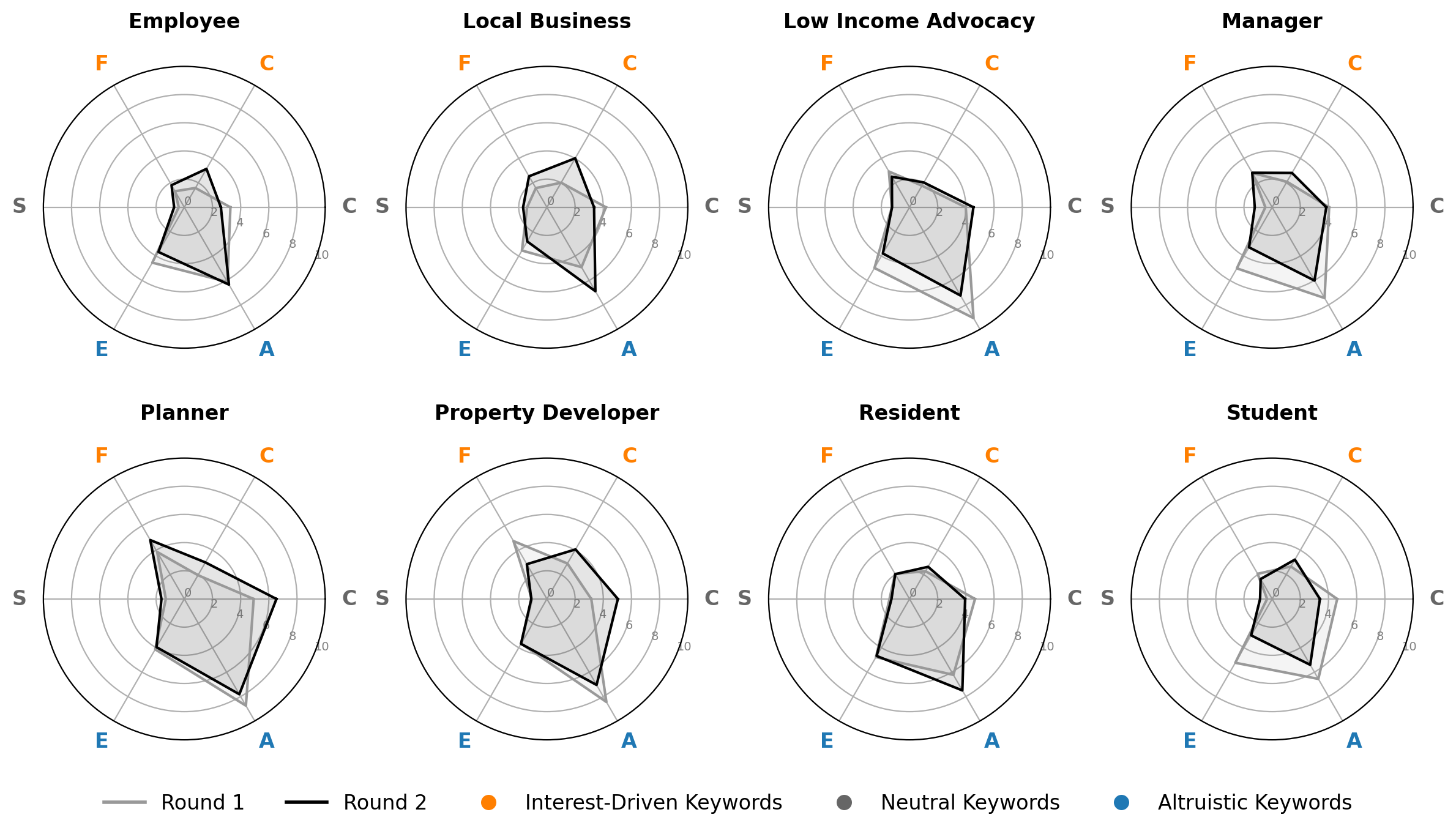}
    \caption{Keyword Frequency of Agent Outputs Influenced by Communication}
    \label{fig:Keyword Frequency of Agent Outputs Influenced by Communication}
\end{figure}

Figure 3 reveals how stakeholders’ reasoning shifted between Round 1 and Round 2. Round 2 took place after each agent had received a broadcast of all stakeholders’ original opinions from Round 1. Overall, agents’ responses became more aligned and focused in Round 2, likely reflecting a convergence of perspectives due to shared information. We also observe that altruistic keywords remain prominent in roles like the Planner and Low-Income Advocacy, reflecting a consistent emphasis on social equity based on the agent’s predefined values. Meanwhile, the interest-driven and neutral keywords generally show a slight decrease in frequency in Round 2, implying that the agents’ reasoning moved toward shared community or balanced value due to the additional inputs with altruist agents, reflecting more comprehensive considerations between financial interest and social responsibilities.

Figures 4 and 5 demonstrate the error point plot of the decision to vote for low-income housing versus shopping mall in terms of the score given by each agent with and without multi-agent communication. The horizontal axis represents each agent, and the vertical axis represents the score they rated for each planning scenario after their discussions. All agents participate in the setups of experiments described in Section 2, each tested with and without communication: 1) without opinions and demographic information, 2) without demographic information, and 3) including both opinions and demographic information. Each vertical bar represents the decision outcome from one experiment by one agent, with the midpoint being the mean score and the higher and lower bounds reflecting the standard deviation of the scores. The three vertical bars for each agent in each planning scenario represent the results of the three experiments, respectively. The results suggest that without communication, most agents exhibit a clear and consistent preference, with relatively low variance. However, when multi-agent communication is introduced, the gap between the two proposals widens for many agents, with increased polarization and greater variance in how each agent rates the options.

\begin{figure}
    \centering
    \includegraphics[width=1\linewidth]{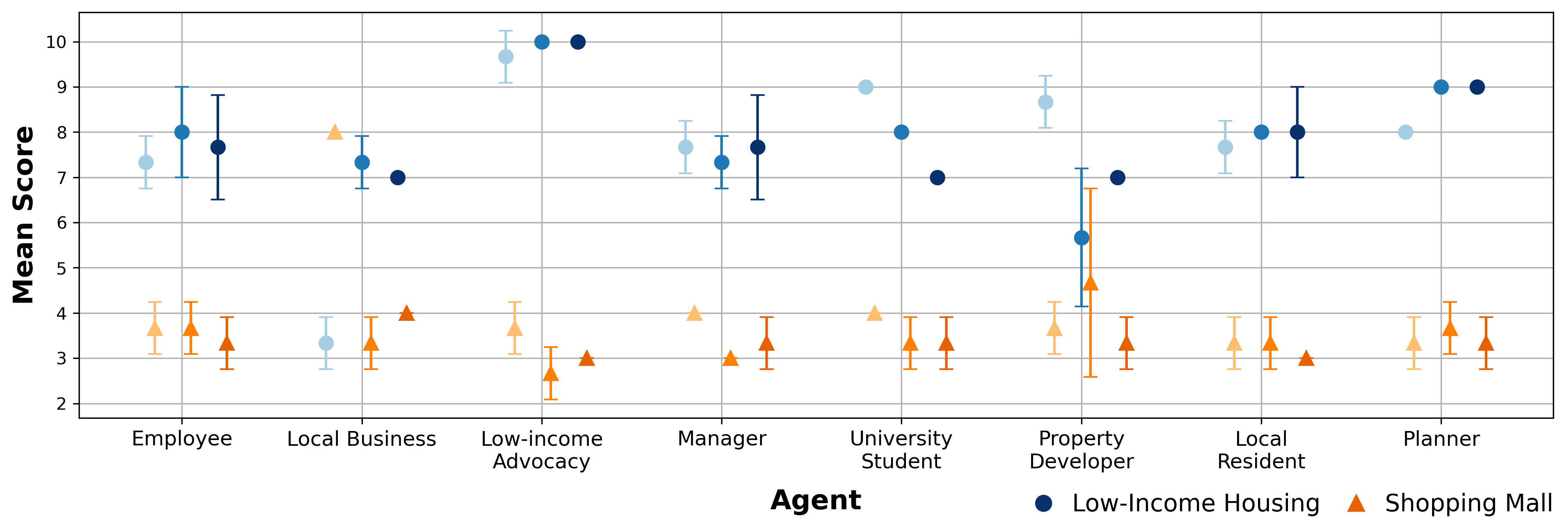}
    \caption{Agent Ratings across Setups without Communication}
    \label{fig:Agent Ratings across Setups without Communication}
\end{figure}

\begin{figure}
    \centering
    \includegraphics[width=1\linewidth]{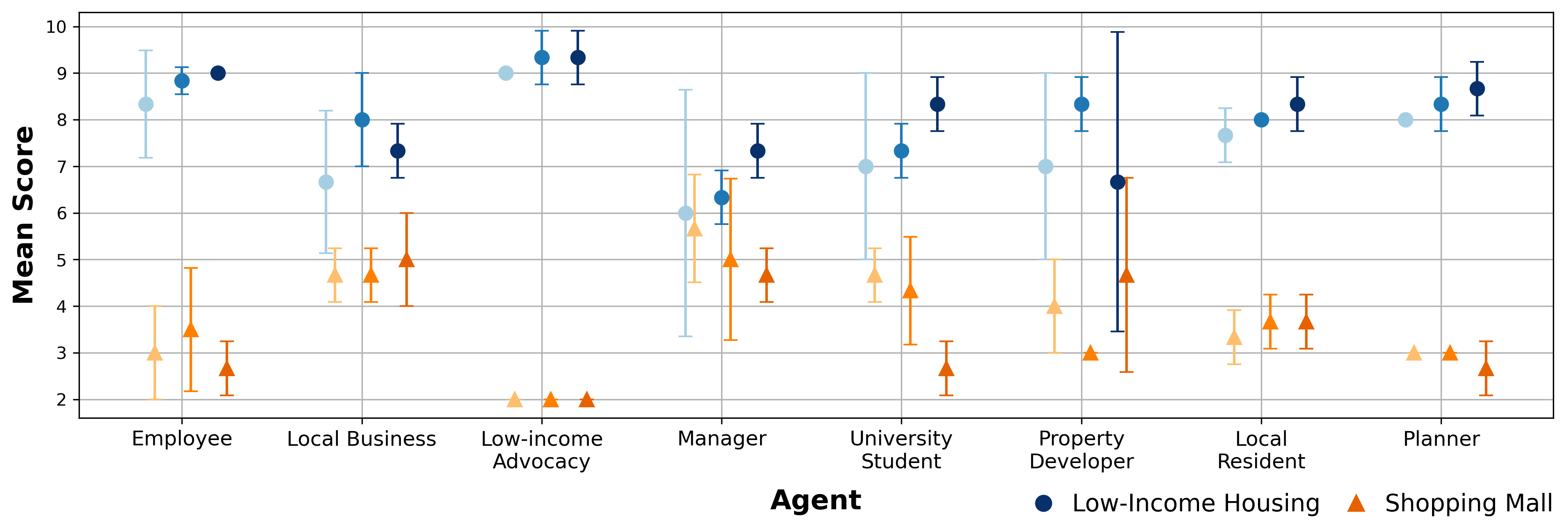}
    \caption{Agent Ratings across Setups with Communication}
    \label{fig:Agent Ratings across Setups with Communication}
\end{figure}

\subsection{Impact of Opinion and Demographic Information}

As shown in Figure 4, in general, most agents tend to provide high ratings for the low-income housing project and low ratings for the shopping mall project. However, considerable variances exist across different agents. Persona provides distinctive choice patterns. The Employee, Low-Income Advocate, Resident, and Urban Planner tend to provide diverging ratings for the two projects. In contrast, the rest of the agents, including the Local Business, Manager, University Student, and Property Developer tend to give the ratings for the low-income housing project closer to that of the shopping mall project. These findings conform with our belief since we would assume that the Low-Income Advocacy would rate higher to the low-income housing project from their perspectives, while the Property Development would rate higher to the shopping mall project for profitability. This could be attributed to the focus of the agent’s personas we extracted.

Adding life values and demographic information helped shape the agents’ opinions. All agents, except for the University Student and Local Business, either changed their decisions or gave higher ratings to the low-income housing project when these factors were included in their profiles. This suggests that the agents' discussions allowed the demand for low-income housing to be heard by multiple stakeholders. In contrast, there was no clear trend in the ratings for the shopping mall project. Furthermore, adding life values and demographic information generally made the agents' ratings more consistent with reduced standard deviation. 

As shown in further keyword analysis in Figure 6, across different setups, the agents’ value orientations remain largely consistent. Adding demographic and opinion data makes certain agents’ value considerations more comprehensive, as reflected in the more clustered radar shapes. For instance, the Low-Income Advocacy agent shows increased frequency of safety and affordability related keywords beyond affordability. The planner agent adds more emphasis on equity while slightly reducing the focus on financial and affordability concerns. This reflects how the introduction of demographic and opinion data allows the agents in the simulation to act more like real-world stakeholder representatives and are better equipped to consider the diverse positions and perspectives of all stakeholders during the decision-making process.

\begin{figure}
    \centering
    \includegraphics[width=1\linewidth]{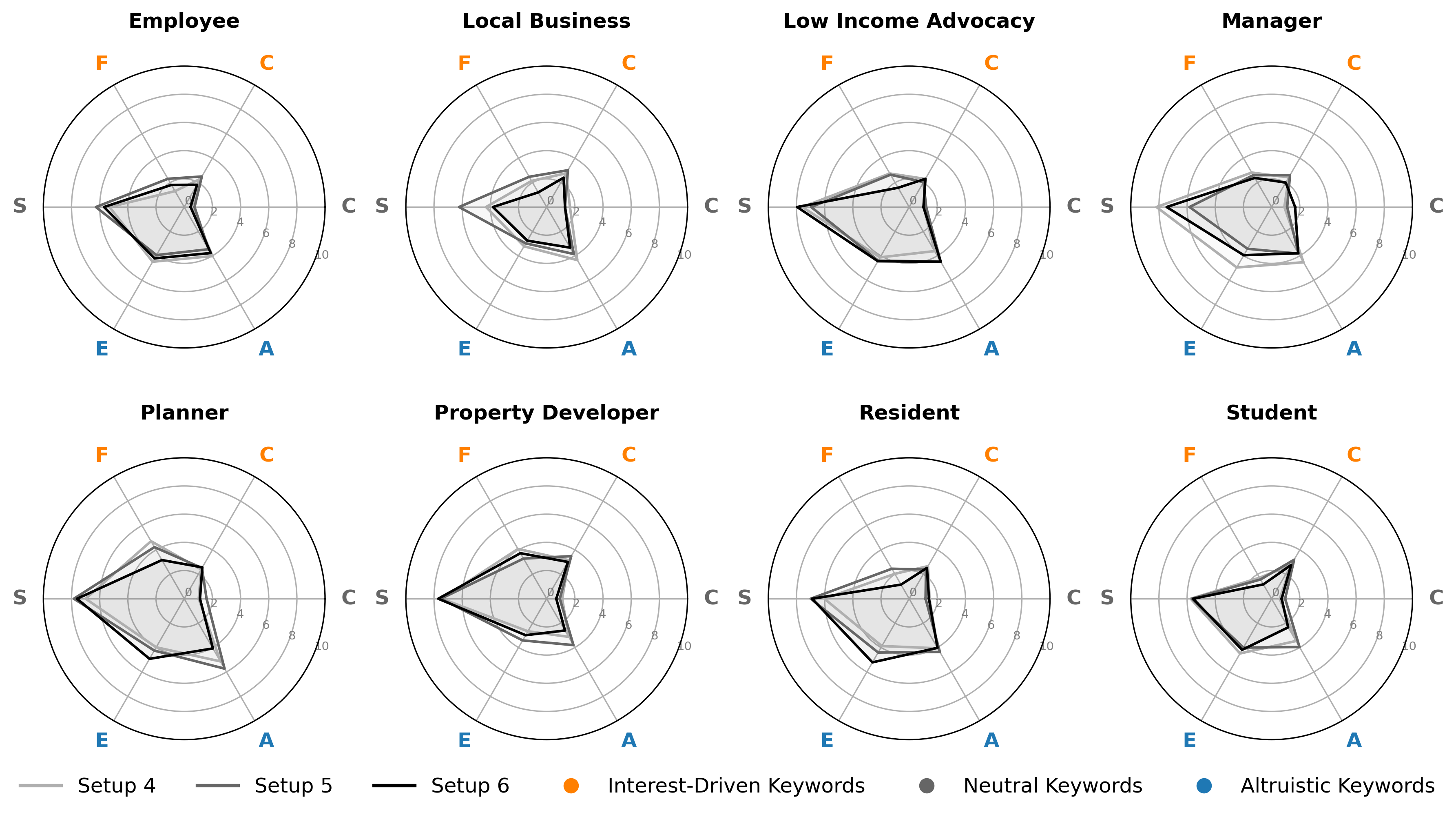}
    \caption{Keyword Frequency of Agent Outputs Based on Experiments.}
    \label{fig:Keyword Frequency of Agent Outputs Based on Experiments.}
\end{figure}

\subsection{Pros and Cons}

One advantage of using simulation as a decision-making auxiliary tool is its efficiency and cost-effectiveness. Gathering feedback from real stakeholders can be time-consuming and expensive, the simulated approach enables iterative testing of various planning proposals in hypothetical scenarios. Additionally, it promotes inclusive voices: unlike real-life processes that often rely on centralized voting, this system enables decentralized decision-making frameworks, ensuring that marginalized populations are represented and their perspectives considered. On the other hand, several drawbacks need to be considered. Unlike real-life stakeholders who may have deeply held beliefs and resist compromise, the generative agents tend to be receptive to others’ opinions, despite prompts emphasizing their own perspectives, which strengthened the echo-chamber effects and convergence on ideas. Additionally, the agents’ built-in moral compass, likely influenced by the underlying LLM models' moral gatekeeper, can lead to an overemphasis on social justice values. Finally, the process of extracting and designing agent personas requires manual interpretation of real-world data, which can inadvertently introduce researcher biases into the simulation.

\subsection{Ethical Risks and Recommendations} 

Several ethical concerns arise in the design of this system. First, if the agent design does not adequately represent the diverse views of all stakeholders, some groups might be misrepresented, leading to decisions that fail to meet the needs of the entire community. Biases in the underlying data can also be amplified by the algorithms, potentially reinforcing existing social inequalities. While including demographic details can improve the comprehensiveness of the agents' responses, it further heightens privacy risks associated with sensitive demographic information.

To address these potential ethical risks, several strategies are needed.  Establishing clear guidelines is essential to ensure the inclusion of diverse populations. It is also important to carefully design and evaluate the entire data collection and processing pipeline, including LLM model selection, interviewee recruitment, persona and agent design, and prompt development, to identify and mitigate potential biases. Additionally, involving external third audit parties can help minimize biases and ensure transparency.

\subsection{Future works} 
Due to computational resource constraints, we were limited to a small set of agents in this experiment. Conducting similar experiments with a broader range of stakeholders and agents representing diverse demographic backgrounds would yield richer insights. Incorporating knowledge from negotiation studies and decision-making frameworks, such as game theory, could also enhance the system’s explanatory power. Finally, exploring alternative prompting strategies and discussion frameworks, including hierarchical power dynamics, may help generate more stable and passionate opinions.

\section{Conclusion}
This study explored the use of a multi-generative agent decision-making system in an urban planning scenario involving diverse stakeholders. We found that multi-agent communication enhances the quality of the agents’ arguments, supporting richer, more innovative, and inclusive reasoning. Integrating demographic and life-value data further improves the diversity and stability of agent outputs. These findings provide a predictive framework for decision-makers to anticipate stakeholder reactions using a simulation-based approach, thereby fostering more equitable and cost-effective urban planning decisions.

\section*{Acknowledgement}

We thank Professors Manish Raghavan and Ashia Wilson (MIT EECS) for their guidance and feedback, as well as Professor Kairos Shen and Mr. John Attanucci (MIT DUSP), and all interview participants for their input in the early stages of this work.

\bibliographystyle{unsrtnat}
\bibliography{references}

@article{park2023generative,
  title     = {Generative Agents: Interactive Simulacra of Human Behavior},
  author    = {Park, Joon Sung and O’Brien, Joseph C. and Cai, Carrie J. and Morris, Meredith Ringel and Liang, Percy and Bernstein, Michael S.},
  journal   = {arXiv preprint arXiv:2304.03442},
  year      = {2023},
  archivePrefix = {arXiv},
  primaryClass  = {cs.AI},
  doi       = {10.48550/arXiv.2304.03442}
}

@article{ren2024emergence,
  title     = {Emergence of Social Norms in Generative Agent Societies: Principles and Architecture},
  author    = {Ren, Shun and Cui, Zhi and Song, Rui and Wang, Zhi and Hu, Shiwei},
  journal   = {arXiv preprint arXiv:2403.08251},
  year      = {2024},
  archivePrefix = {arXiv},
  primaryClass  = {cs.AI},
  doi       = {10.48550/arXiv.2403.08251}
}

@article{dai2024artificial,
  title     = {Artificial Leviathan: Exploring Social Evolution of LLM Agents through the Lens of Hobbesian Social Contract Theory},
  author    = {Dai, Guangyu and Zhang, Wen and Li, Junjie and Yang, Shuo and Lbe, Clement O. and Rao, Shriram and Caetano, Antonio and Sra, Misha},
  journal   = {arXiv preprint arXiv:2406.14373},
  year      = {2024},
  archivePrefix = {arXiv},
  primaryClass  = {cs.AI},
  doi       = {10.48550/arXiv.2406.14373}
}

@article{costabile2025factchecking,
  title     = {Assessing the Potential of Generative Agents in Crowdsourced Fact-Checking},
  author    = {Costabile, Luigi and Orlando, Gianluca M. and Gatta, Vincenzo L. and Moscato, Vincenzo},
  journal   = {arXiv preprint arXiv:2504.19940},
  year      = {2025},
  archivePrefix = {arXiv},
  primaryClass  = {cs.AI},
  doi       = {10.48550/arXiv.2504.19940}
}

@article{bakhtin2022diplomacy,
  title     = {Human-Level Play in the Game of Diplomacy by Combining Language Models with Strategic Reasoning},
  author    = {Bakhtin, Anton and Brown, Noam and Dinan, Emily and Farina, Gabriele and Flaherty, Caitlin and Fried, Daniel and Goff, Alex and Gray, John and Hu, Hengyuan and Jacob, Aaron P. and others},
  journal   = {Science},
  volume    = {378},
  number    = {6624},
  pages     = {1067--1074},
  year      = {2022},
  doi       = {10.1126/science.ade9097}
}

@article{park2024thousand,
  title     = {Generative Agent Simulations of 1,000 People},
  author    = {Park, Joon Sung and Zou, Catherine Q. and Shaw, Aaron and Hill, Benjamin M. and Cai, Carrie J. and Morris, Meredith Ringel and Willer, Robb and Liang, Percy and Bernstein, Michael S.},
  journal   = {arXiv preprint arXiv:2411.10109},
  year      = {2024},
  archivePrefix = {arXiv},
  primaryClass  = {cs.AI},
  doi       = {10.48550/arXiv.2411.10109}
}

@article{chopra2024limits,
  title     = {On the Limits of Agency in Agent-Based Models},
  author    = {Chopra, Abhishek and Kumar, Shubham and Giray-Kuru, Nazli and Raskar, Ramesh and Quera-Bofarull, Arnau},
  journal   = {arXiv preprint arXiv:2409.10568},
  year      = {2024},
  archivePrefix = {arXiv},
  primaryClass  = {cs.AI},
  doi       = {10.48550/arXiv.2409.10568}
}

@article{hou2025vaccine,
  title     = {Can a Society of Generative Agents Simulate Human Behavior and Inform Public Health Policy? A Case Study on Vaccine Hesitancy},
  author    = {Hou, Andrew B. and Du, Haoyi and Wang, Yixuan and Zhang, Junjie and Wang, Zhen and Liang, Paul P. and Khashabi, Daniel and Gardner, Lauren and He, Tian},
  journal   = {arXiv preprint arXiv:2503.09639},
  year      = {2025},
  archivePrefix = {arXiv},
  primaryClass  = {cs.AI},
  doi       = {10.48550/arXiv.2503.09639}
}

@article{piao2025polarization,
  title     = {Emergence of Human-Like Polarization among Large Language Model Agents},
  author    = {Piao, Jiarui and Lu, Zhaoyang and Gao, Chao and Xu, Feiyang and Hu, Qi and Santos, Francisco P. and Li, Yunzhi and Evans, James},
  journal   = {arXiv preprint arXiv:2501.05171},
  year      = {2025},
  archivePrefix = {arXiv},
  primaryClass  = {cs.AI},
  doi       = {10.48550/arXiv.2501.05171}
}

@article{ashery2025conventions,
  title     = {Emergent Social Conventions and Collective Bias in LLM Populations},
  author    = {Ashery, A. F. and Aiello, Luca Maria and Baronchelli, Andrea},
  journal   = {Science Advances},
  volume    = {11},
  number    = {20},
  pages     = {eadu9368},
  year      = {2025},
  doi       = {10.1126/sciadv.adu9368}
}

@article{ranjan2025fairness,
  title     = {Fairness in Agentic AI: A Unified Framework for Ethical and Equitable Multi-Agent Systems},
  author    = {Ranjan, Rakesh and Gupta, Shubham and Singh, S. N.},
  journal   = {arXiv preprint arXiv:2502.07254},
  year      = {2025},
  archivePrefix = {arXiv},
  primaryClass  = {cs.AI},
  doi       = {10.48550/arXiv.2502.07254}
}

@inproceedings{ondula2024sentimental,
  title     = {Sentimental Agents: Exploring Deliberation, Cognitive Biases, and Decision-Making in LLM-Based Multi-Agent Systems},
  author    = {Ondula, Elizabeth and Orner, David and Mumero, Nicholas and Rusti, Chiara},
  booktitle = {Proceedings of the Fourth Workshop on Knowledge-Infused Learning},
  year      = {2024},
  address   = {Vienna},
  url       = {https://openreview.net/forum?id=izfJXk4wGz}
}

@article{bai2024fairmonitor,
  title     = {FairMonitor: A Dual-Framework for Detecting Stereotypes and Biases in Large Language Models},
  author    = {Bai, Yuxuan and Zhao, Jing and Shi, Jing and Xie, Zihan and Wu, Xiaoyan and He, Lei},
  journal   = {arXiv preprint arXiv:2405.03098},
  year      = {2024},
  archivePrefix = {arXiv},
  primaryClass  = {cs.AI},
  doi       = {10.48550/arXiv.2405.03098}
}

@techreport{mit2021volpe,
  title     = {MIT Volpe Final Development Plan, Volume 1},
  author    = {{Massachusetts Institute of Technology}},
  institution = {City of Cambridge Planning Board},
  number    = {PB-368},
  year      = {2021},
  url       = {https://www.cambridgema.gov/-/media/Files/CDD/ZoningDevel/SpecialPermits/sp368/sp368_appnarrative_20210603.pdf}
}

@article{wu2023autogen,
  title     = {AutoGen: Enabling Next-Gen LLM Applications via Multi-Agent Conversation},
  author    = {Wu, Qingyun and Bansal, Gagan and Zhang, Jing and Wu, Yao and Li, Beibin and Zhu, Erheng and Jiang, Linxi and Zhang, Xinyu and Zhang, Shuo and Liu, Jiahui and others},
  journal   = {arXiv preprint arXiv:2308.08155},
  year      = {2023},
  archivePrefix = {arXiv},
  primaryClass  = {cs.AI},
  doi       = {10.48550/arXiv.2308.08155}
}

\end{document}